\documentclass[a4paper]{jpconf}
\usepackage{amsmath,amsfonts,amssymb,graphicx,color,
graphicx,subfigure,hyperref,cite}

\begin{document}
\title{Disappointing model for ultrahigh-energy cosmic rays}

\author{R Aloisio$^1$, V Berezinsky$^{1,2}$, and A Gazizov$^{2,3}$}

\address{$^1$ INFN, National Gran Sasso Laboratory, I-67010 Assergi (AQ), Italy}
\address{$^2$ Gran Sasso Astroparticle Center, I-67010 Assergi (AQ), Italy}
\address{$^3$ Institute of Physics of NASB, 68 Independence Avenue, BY-22072 Minsk, Belarus}

\ead{askhat.gazizov@lngs.infn.it}

\begin{abstract}
Data of Pierre Auger Observatory show a proton-dominated chemical composition of ultrahigh-energy cosmic rays spectrum at ($1 - 3$)~EeV and a steadily heavier composition with energy increasing. In order to explain this feature we assume that ($1 - 3$)~EeV protons are extragalactic and derive their maximum acceleration energy,  $E_p^{\rm max} \simeq 4$~EeV, compatible with both the spectrum and the composition. We also assume the rigidity-dependent acceleration mechanism of heavier nuclei, $E_A^{\rm max} = Z \times E_p^{\rm max}$. The proposed model has rather disappointing consequences: i) no pion photo-production on CMB photons in extragalactic space and hence ii) no high-energy cosmogenic neutrino fluxes; iii) no GZK-cutoff in the spectrum; iv) no correlation with nearby sources due to nuclei deflection in the galactic magnetic fields up to highest energies. 
\end{abstract}

Spectra and chemical compositions of ultrahigh-energy ($E \gtrsim 1$ EeV) cosmic rays (UHECR) measured by two largest detectors, High Resolution Fly's Eye (HiRes) \cite{Abbasi:2007sv} and Pierre Auger Observatory (PAO) \cite{Abraham:2008ru}, are significantly different. 

The HiRes data show pure \emph{proton} composition  \cite{sokol-trondheim,Sokolsky:2010kb}, confirming such signatures of their propagation through CMBR as the GZK cutoff \cite{Greisen:1966jv,Zatsepin:1966jv} and the pair-production dip \cite{Berezinsky:1988wi,Berezinsky:2002nc,
Aloisio:2006wv,Berezinsky:2002vt,Berezinsky:2005cq}. 
\begin{figure*}[ht]
\centering
\mbox{\subfigure{\includegraphics[height=50mm,width=65mm]{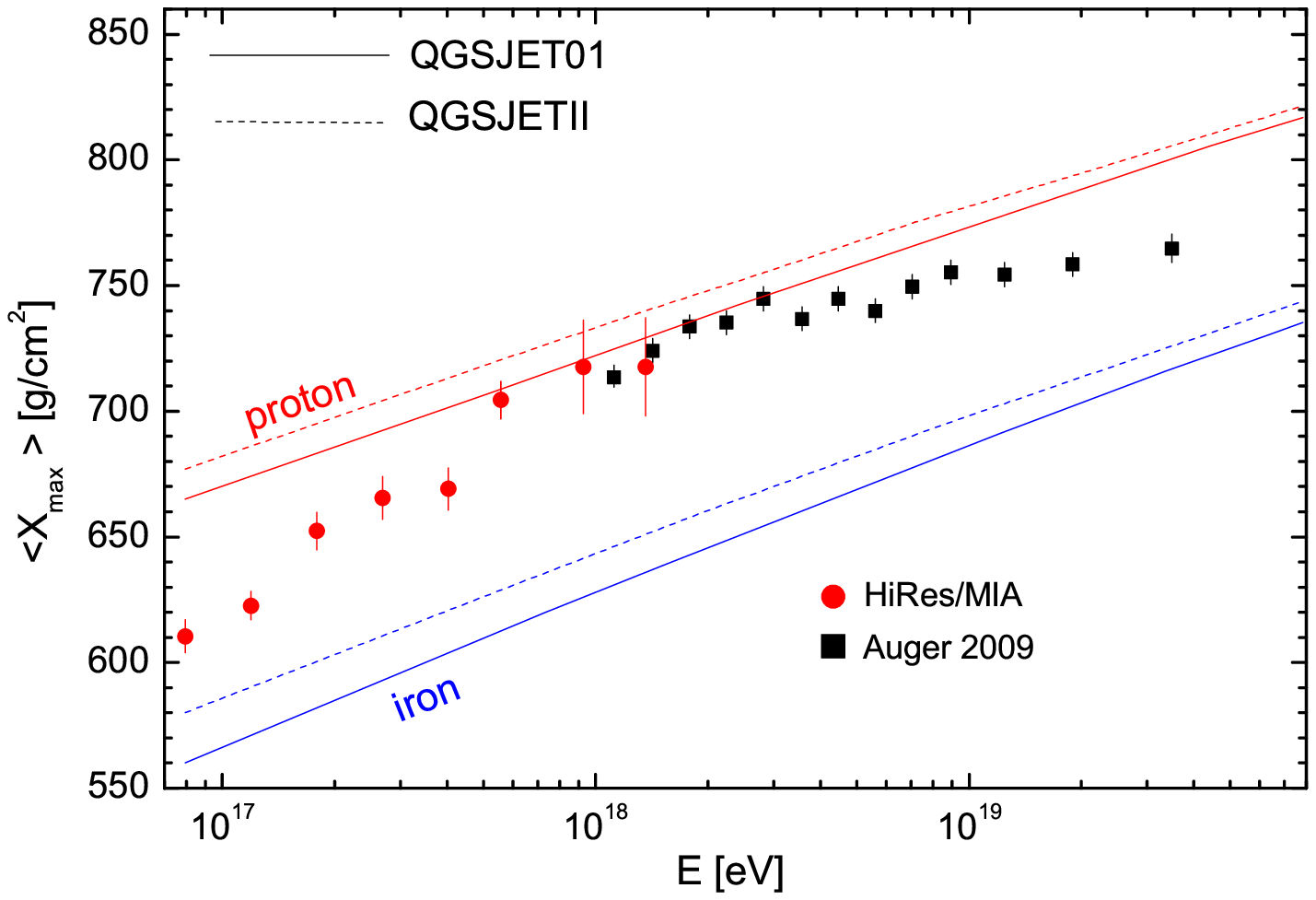}}\quad
\subfigure{\includegraphics[height=51mm]{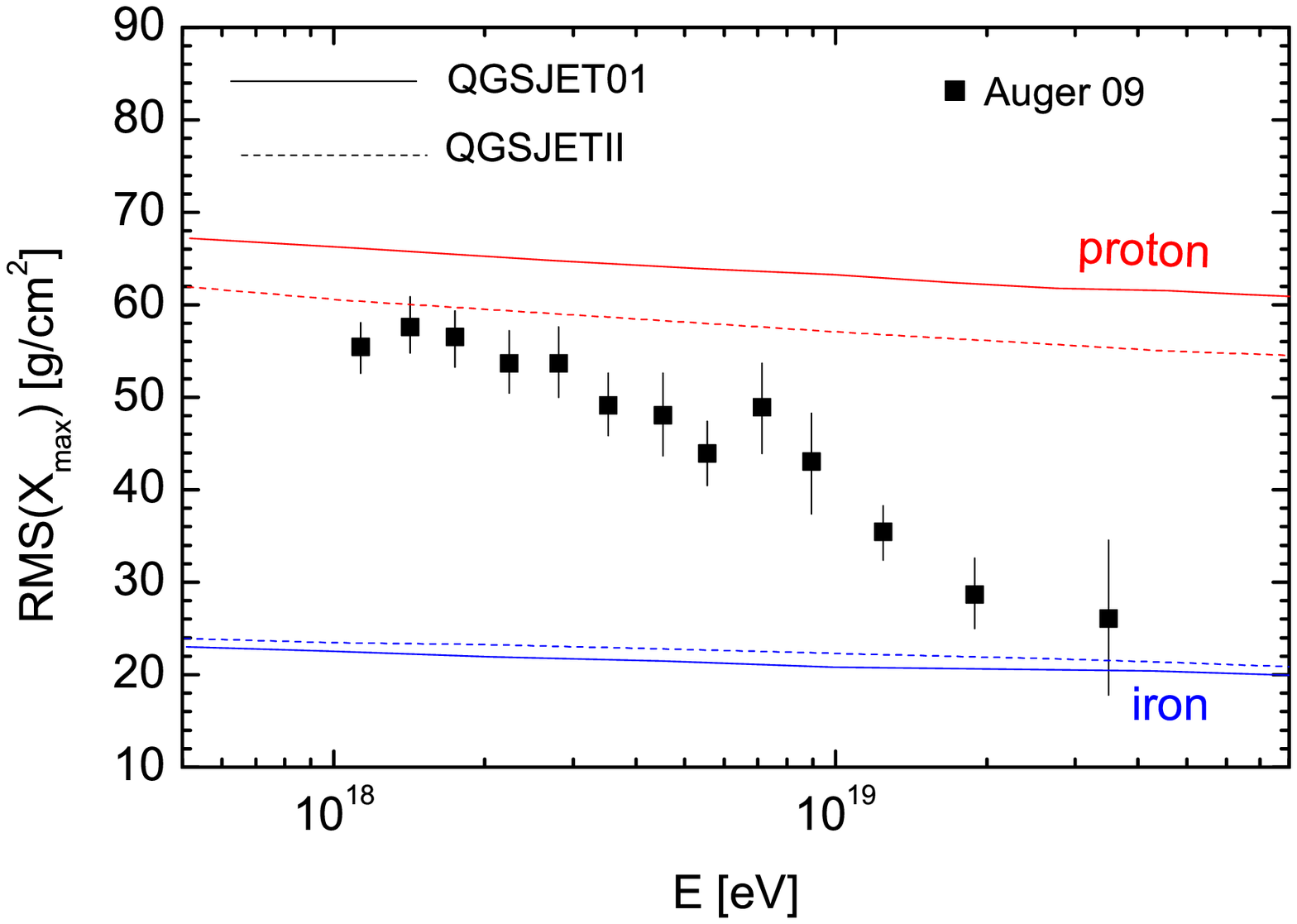}}}
\caption{
PAO data \cite{Auger-data,Abraham:2010yv,Bellido} on $X_{\rm max}(E)$ (left panel) and on RMS($X_{\rm max}$) (right panel). Lines for protons and Iron are according to QGSJET model \cite{Ostapchenko:2005nj}. 
} 
\label{Fig3}%
\end{figure*}

The PAO data, on the contrary, strongly favor the nuclei composition getting progressively heavier at $E \simeq (4 - 40)$~EeV. This feature, in terms of energy dependence of EAS development maximum in atmosphere, $X_{\rm max}(E)$, and r.m.s.\ of this observable, RMS($X_{\rm max}$), is clearly seen in Fig.~\ref{Fig3}. The data also suggest that the nucleus charge number $Z$ changes smoothly in sources.

Here we demonstrate that the simple, but disappointing for future experiments, model \cite{Aloisio:2009sj} can naturally explain both energy spectrum and mass composition observed by the PAO. 

The basic assumption of the model is the \emph{proton composition} of UHECR spectrum at $E \simeq (1 - 3)$~EeV, the feature supported both by PAO and HiRes. Two more assumptions are that these protons are \emph{extragalactic} and that acceleration of primary nuclei in sources is \emph{rigidity-dependent}, i.e.\ that  
$E_{\rm max}^{\rm acc} = Z \times E_0$, where $E_0$ is a universal energy to be
determined from data; $Z$ is a nucleus charge number. 

\begin{figure*}[ht]
\centering
\mbox{
\subfigure{\includegraphics[height=50mm,width=65mm]{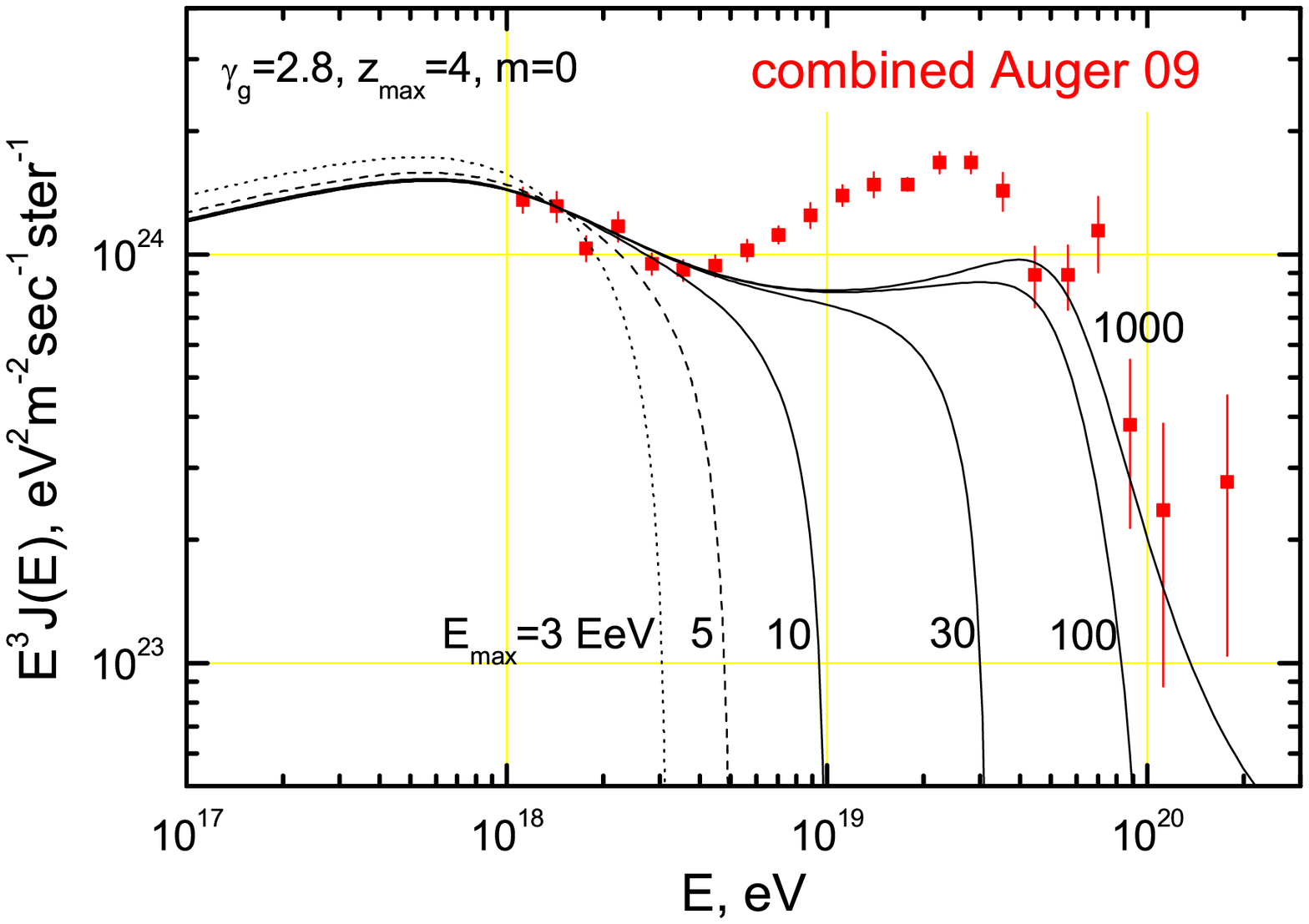}}
\quad 
\subfigure{\includegraphics[height=50mm,width=65mm]{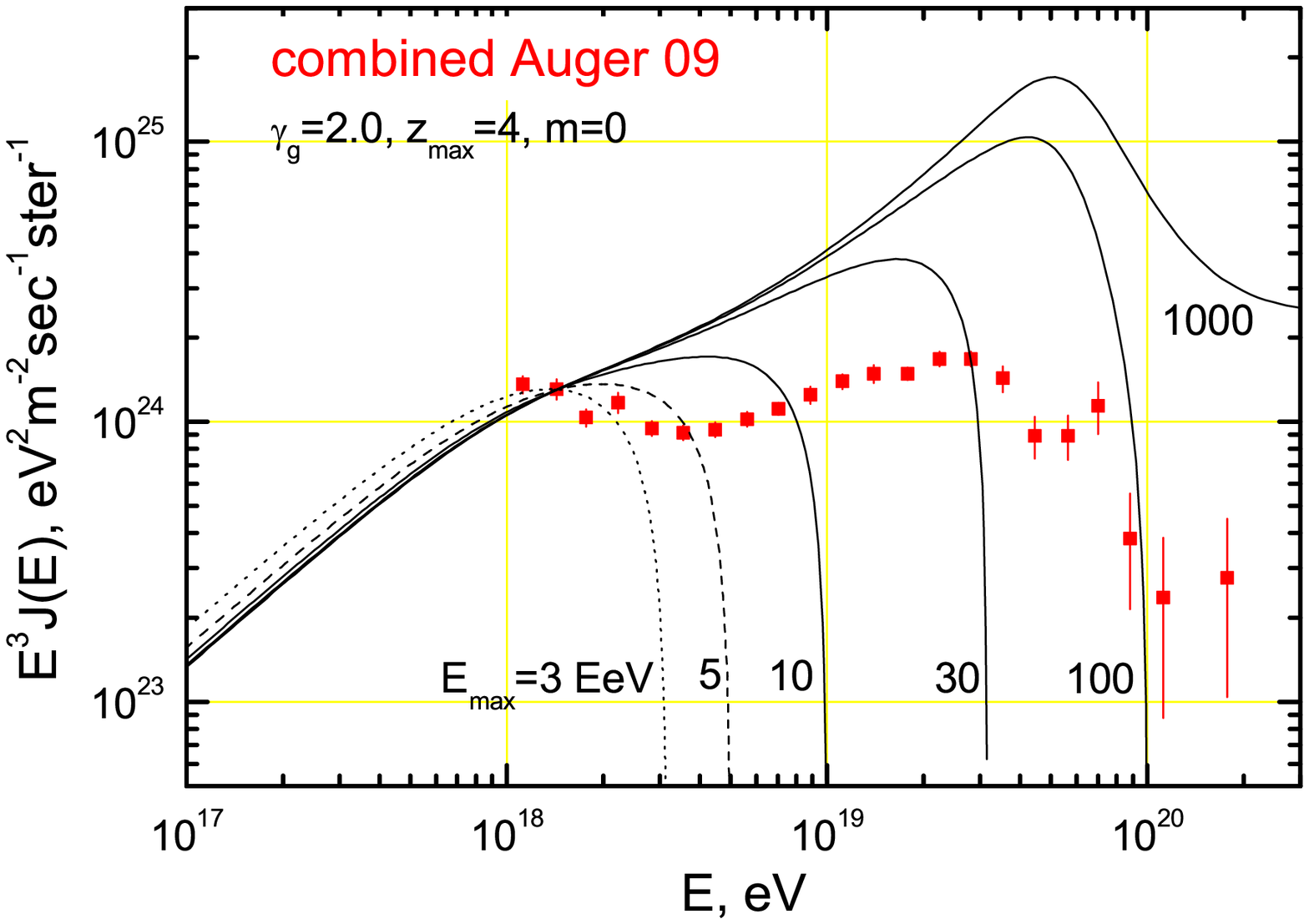}}
}
\caption{
Calculated proton spectra compared to the combined PAO 
spectrum for different $E_p^{\rm max}$. Extreme cases $\gamma_g=2.8$ 
and 2.0 are shown in the left and right panels, respectively. 
} 
\label{Fig4}%
\end{figure*}

In order to determine the maximum acceleration energy of protons, 
$E_p^{\rm max}=E_0$, let us calculate the extragalactic diffuse 
proton flux, assuming the power-law generation spectrum 
$Q_g(E) \propto E^{-\gamma_g}$ with $E_{\rm max}=E_0$, and normalize 
it by the PAO flux at ($1 - 3$)~EeV. Varying $\gamma_g$ in the range $2.0 - 2.8$, the maximum value of $E_0$ allowed by the PAO mass composition (see Fig.~\ref{Fig3}) and energy spectrum (see Fig.~\ref{Fig4}) may be obtained. 

In our calculations a homogeneous distribution of sources with no cosmological evolution ($m=0$) was assumed; the highest redshift of sources $z_{\rm max}=4$. As a criterion of contradiction an excess of calculated proton flux at $E \sim (4 - 5)$~EeV was chosen. The contradiction has different character for different values of $\gamma_g$. 

For steep source generation functions with $\gamma_g \simeq 
2.6 - 2.7$ the shape and flux of the PAO spectrum may be described 
by $E_p^{\rm max} \sim 10^{20} - 10^{21}$~eV; the contradiction occurs 
only in data on mass composition. The extreme case, given by 
$\gamma_g=2.8$, is displayed in the left panel of Fig.~\ref{Fig4}. 

For flat generation spectra (see the extreme case of $\gamma_g=2.0$ in the right panel of Fig.~\ref{Fig4}) the contradiction is very pronounced. For $E_p^{\rm max}= 5$~EeV the calculated proton 
flux exceeds the observed one even at $E \approx 2$~EeV.
 
It is clear that with some redundancy $E_p^{\rm max} \simeq  
(4 - 6)$~EeV for all $2.0 \lesssim \gamma_g \lesssim 2.8$. 

\begin{figure*}[t]
\centering
\mbox{\subfigure{\includegraphics[height=50mm,width=65mm]{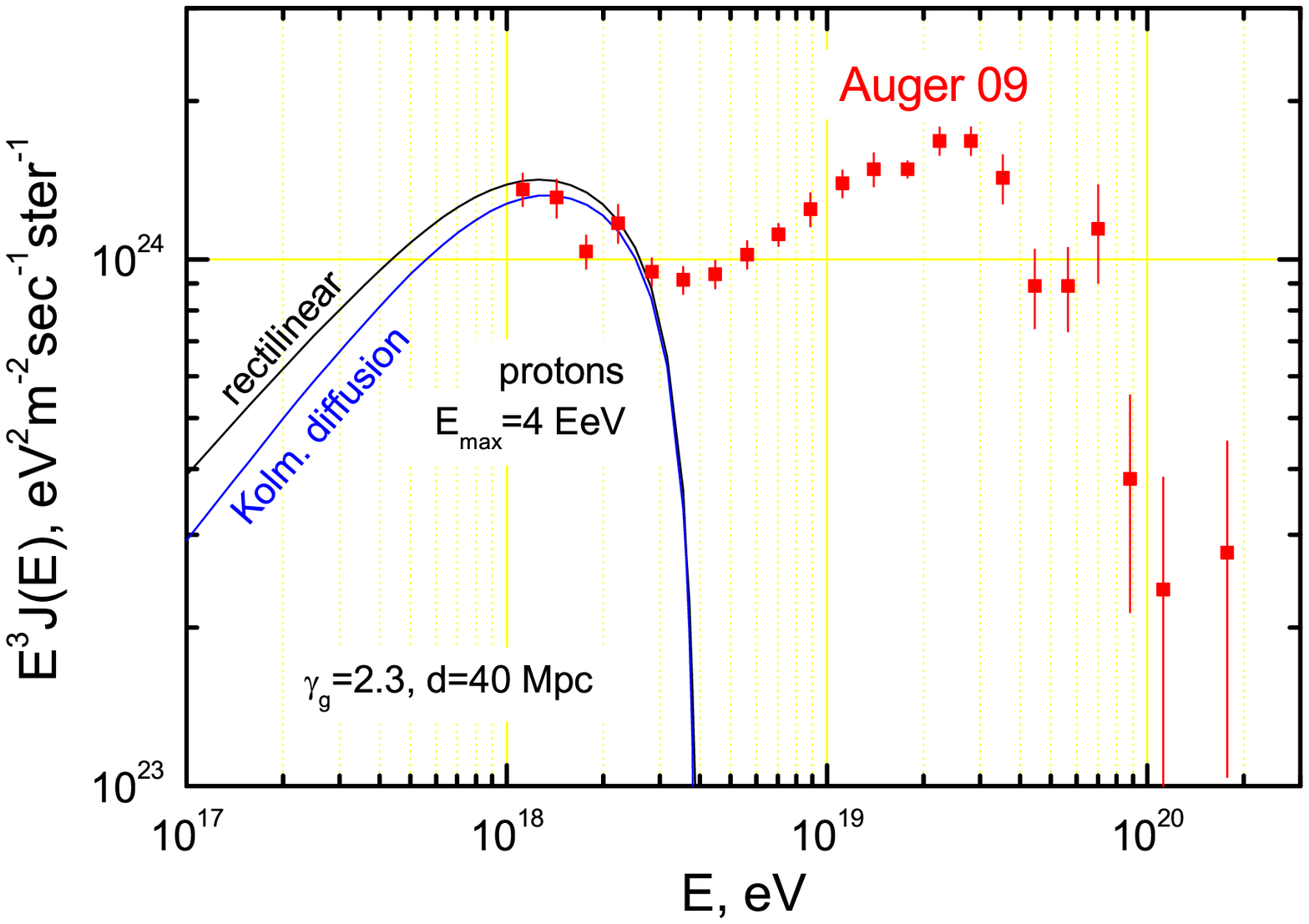}}\quad
\subfigure{\includegraphics[height=50mm,width=65mm]{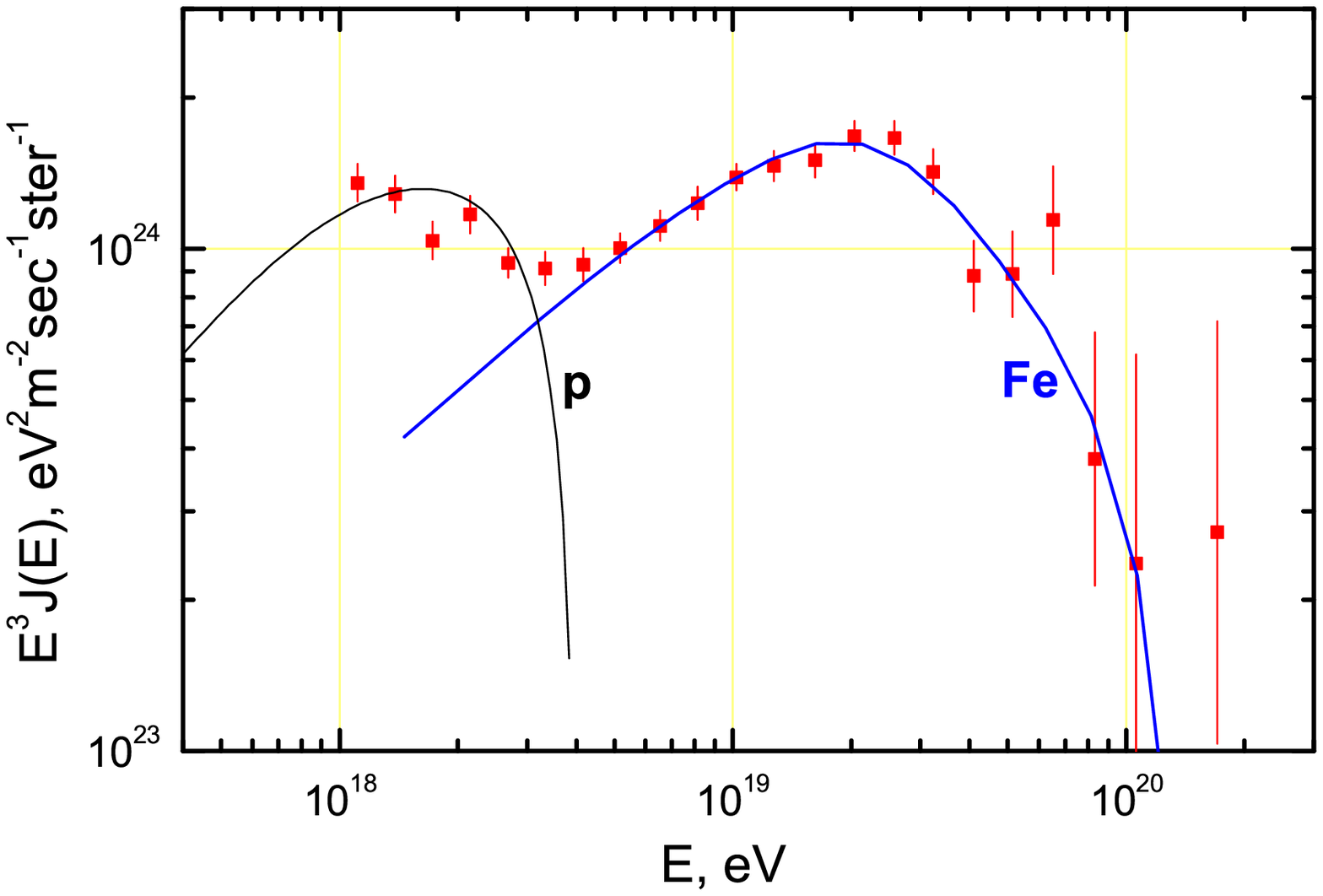}}}
\caption{
{\em Left panel:} Comparison of calculated proton spectra with the combined PAO spectrum for $\gamma_g=2.3$ and diffusive proton propagation. The cutoff at $E_p^{\rm max} = 4$~EeV is needed to avoid 
the contradiction with data at $E> 3~$EeV. {\em Right panel:} The energy spectrum in two-component model with protons and Iron nuclei with $\gamma_g=2.0$ and $E_{\max}=4 \times Z$~EeV. The Iron nuclei spectrum is calculated for homogeneous distribution of the sources.
} 
\label{Fig5}%
\end{figure*}

An influence of possible intergalactic magnetic fields on proton spectrum calculated in a diffusive model is shown in the left panel of Fig.~\ref{Fig5}. Here $\gamma_g =2.3$, which might be the case for acceleration by relativistic shocks. The Kolmogorov diffusion in turbulent magnetic field with basic scales $(B_c,l_c)=(1 \mbox{ nG, } 1\mbox{ Mpc})$ was assumed (see \cite{Berezinsky:2005fa,Aloisio:2008tx}) and distances between sources were $d \simeq 40$~Mpc. The analysis of proton maximum energy of acceleration gives again $E_0=E_p^{\rm max}=4$~EeV, in a rough agreement with the analysis made for homogeneous distribution of sources. The account for diffusion brings to the flattening of the proton spectrum at $E \lesssim 1$~EeV, seen in Fig.~\ref{Fig5} as a 'diffusive cutoff', which provides a transition from the steep galactic spectrum, most probably composed of Iron, to the flat spectrum of extragalactic protons. 

The basic feature of the PAO mass composition, the progressively heavier composition with energy increasing, is 
guaranteed in our model by the rigidity-dependent maximum energy of 
acceleration: at energy higher than $Z \times E_p^{\max}$ nuclei with charge $Z' < Z$ disappear, while heavier nuclei with larger $Z$ survive. Starting from $E_p^{\rm max} \sim (4 - 6)$~EeV, the higher energies are accessible only for nuclei with progressively larger values of $Z$. 

Let us now consider a two-component model, with only protons and Iron nuclei being produced in sources with generation index $\gamma_g=2.0$ and the maximum acceleration energy $E_{\max}=4 \times Z$~EeV, shown in the right panel of Fig.~\ref{Fig5}. The primary Iron nuclei spectrum is calculated as in  \cite{Aloisio:2008pp,Aloisio:2010he} for homogeneous distribution of sources. One may notice that the calculated spectrum of Iron describes well the cutoff in the PAO spectrum. This steepening is caused by the photo-disintegration of Iron nuclei.

\begin{figure*}[ht]
\begin{minipage}[b]{0.49\linewidth}
\centering
\mbox{\includegraphics[height=50mm,width=65mm]{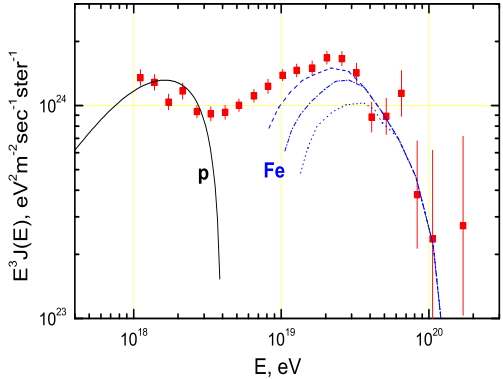}}
\caption{
As in the right panel of Fig. 3, but with the `'diffusion cutoff' 
introduced for three different sets  of parameters $B_c, l_c, d$. The
gap between 3~EeV and $E_{\rm cut}$ (beginning of 'diffusive cutoff')
is expected to be filled by intermediate nuclei with $2 \leq Z \leq 25$. 
} 
\label{Fig6}%
\end{minipage}
\hspace{2 mm}
\begin{minipage}[b]{0.48\linewidth}
with the proton composition (the 6th low-energy bin of the PAO data in Fig.~\ref{Fig3}). If this energy increases, $E_p^{\max}$ increases, too. The model collapses when the allowed $E_p^{\max}$ reaches e.g.\ ($50 - 100$)~EeV. 
 
\hspace{3mm} Another case is given by the mass composition beinglight nuclei starting right from 1~EeV \cite{Aloisio:2008tx}. The cosmological evolution of sources are not included in our calculations; since this effect slightly decreases 
$E_p^{\rm max}$, it is not needed to be taken into account. In principle, it is also possible that the EeV protons detected by PAO are secondary ones, i.e.\ those produced in photo-dissociation of primary nuclei in collisions with CMBR and extragalactic IR/UV photons.
However, in fact, as it was demonstrated in \cite{Allard:2008gj,Aloisio:2010he}, the flux of secondary protons in the EeV range is always smaller than the sum of primary and secondary nuclei fluxes. 
\end{minipage}
\end{figure*}

To agree with the mass composition of PAO, the Iron spectrum  
in Fig.~\ref{Fig5} must have a low-energy cutoff at $E \lesssim 
(20 - 30)$~EeV. Most naturally it is produced as a 'diffusive cutoff'  
which appears in models with lattice-located sources due to
{\em magnetic horizon} \cite{Aloisio:2004jda}. Such cutoffs are shown
in Fig.~\ref{Fig6} for three different sets of parameters $B_c, l_c, d$. The beginning of this cutoff $E_c$ for Iron nuclei is $Z=26$ times higher than for protons, i.e.\ $E_c \approx 2.6 \times 10^{19}$~eV, which has a reasonable physical meaning.  The gap between $2$~EeV and $26$~EeV is expected to be filled by intermediate nuclei. To provide a smooth RMS($X_{\rm max}$) curve seen in Fig.~\ref{Fig3}, there are many free parameters, e.g.\ arbitrary fractions of nuclei accelerated in distant sources. 

The predictions of our model are very disappointing for the future detectors. Really, the maximum acceleration energy $E_{\rm max} \sim (100 - 200)$ EeV for Iron nuclei implies the energy per nucleon $E_p < E_{\rm max}/A \sim (2 - 4) \mbox{ EeV}$, well below the GZK cutoff for epochs with $z \lesssim 15$. Therefore, practically no cosmogenic neutrinos can be produced in collisions of protons and nuclei with CMB photons. 
Correlation with UHECR sources also is absent due to deflection of nuclei in the galactic magnetic fields. The lack of correlation in the model is strengthened by the dependence of the maximum energy on $Z$. 

The signatures of the 'disappointing model' for the PAO detector are the mass-energy relation, already seen in the elongation curve $X_{\rm max}(E)$, and transition from galactic to extragalactic cosmic rays  
below the characteristic energy $E_c \sim 1$~EeV. 

There are some uncertainties in the model presented above. The most important one relates to estimates of $E_p^{\max}$. It is determined by the lowest energy where PAO data become inconsistent

\ack
The work of A. Gazizov was supported by a contract with Gran Sasso Center for Astroparticle Physics (CFA) funded by European Union and Regione Abruzzo under the contract P.O. FSE Abruzzo 2007-2013, Ob.\ CRO.

\section*{References}
\providecommand{\newblock}{}

\end{document}